\documentstyle[12pt,axodraw,epsfig]{article}
\include{dspace12}
\def\baselinestretch{1.3}
\newcommand{\ba}{\begin{array}}
\newcommand{\ea}{\end{array}}
\newcommand{\bd}{\begin{displaymath}}
\newcommand{\ed}{\end{displaymath}}
\newcommand{\be}{\begin{equation}}
\newcommand{\ee}{\end{equation}}
\newcommand{\bea}{\begin{eqnarray}}
\newcommand{\eea}{\end{eqnarray}}

\def\q2 {q^2}

\def\slash {\!\!\!\!/}
\def\Tilde {\widetilde}
\newcommand{\lsim}
{{\;\raise0.3ex\hbox{$<$\kern-0.75em\raise-1.1ex\hbox{$\sim$}}\;}}

\def\neu{\Tilde\chi^0}
\def\slep {\tilde{\ell}}

\def\tanb {\tan \beta}
\begin{document}

\begin{flushright}
HRI-P-04-08-002\\
hep-ph/0408296
\end{flushright}

\begin{center}
{\Large\bf R-parity violation in split supersymmetry}\\[20mm]
Sudhir Kumar Gupta$^{a,}$\footnote{E-mail: guptask@mri.ernet.in}, 
Partha Konar$^{b,}$\footnote{E-mail: konar@theory.tifr.res.in} and
Biswarup Mukhopadhyaya$^{a,}$\footnote{E-mail: biswarup@mri.ernet.in}\\

$^a${\em Harish-Chandra Research Institute, Chhatnag Road, Jhusi,\\
Allahabad - 211 019, India}\\

$^b${\em Department of Theoretical Physics,
Tata Institute of Fundamental Research,\\
Homi Bhabha Road, Mumbai 400005, India}
\end{center}
 \vskip 20 mm

\begin{abstract}
In the recently proposed `split supersymmetry' scenario, the squark
and slepton masses are allowed to be at a high scale while the gauginos
and Higgsinos are within a TeV. We show that in a theory with broken
R-parity, the parameter space of such a scenario allows a situation where
the lightest neutralino is still stable on the cosmological scale and 
can be a dark matter candidate. We also 
separate the cases where (a) it may be invisible but not a dark matter 
candidate, or (b) it may decay showing a displaced vertex. It is also
emphasized how the constraint on the simultaneous violation of baryon and
lepton numbers gets relaxed in this scenario. 
\end{abstract}

\vskip 0.5 true cm

{\bf PACS :}  12.60.Jv, 
	      95.35.+d, 
	      14.80.Ly\\
	
{\bf Keywords :} Split Supersymmetry, Lightest supersymmetric particle, Broken R-Parity.\\

\vskip 1 true cm
\newpage
\setcounter{footnote}{0}
\def\baselinestretch{1.8}

\noindent
The quest for a supersymmetric nature is several decades old
now~\cite{susy}.  Keeping the hierarchy problem in mind, the
proponents of supersymmetry (SUSY), as well as those involved in
devising search strategies for it, have assumed all along that the
SUSY breaking scale is on the order of a TeV, where the masses of all
the superpartners of the standard model particles should lie. However,
some recent attempts have sought to find alternatives to this
assumption.  These resort to a so-called {\em split-supersymmetry}
scenario~\cite{ad}. The main features of this scenario and the
arguments leading to it can be summarised as follows:

\begin{itemize}
\item Although SUSY helps us avoid fine-tuning the Higgs mass, a 
broken SUSY almost invariably leads to a large cosmological constant, 
the escape from which is fine-tuning of a more severe kind~\cite{cc}.

\item If ones appeals to the `landscape scenario' in string 
theory~\cite{lands}, where different choices of the string vacuum give
rise to a very large number of possible universes, then it may not be
too improbable~\cite{prob} that we reside in one of them where the
cosmological constant is small enough to allow galaxy formation.

\item Under such circumstances, the Higgs mass which requires a smaller
degree of fine-tuning than the cosmological constant may also be
probabilistically not too `unnatural'.

\item This frees SUSY (which, among other things, may be part of nature 
as a necessary ingredient of superstrings and may also be helpful in
building Grand Unified Theories (GUT)~\cite{sugut} ) from the
requirement of being broken
within the TeV scale. A higher SUSY breaking scale, shown to be
consistent upto about $10^{13}$ GeV, implies squark and slepton masses
of similar order, thereby avoiding the flavour changing neutral
current problem.

\item Such heavy squarks and sleptons do not affect the unification
of coupling constants since they form complete GUT
multiplets~\cite{gut}. One of the ways to ensure this is to have the
gauginos and Higgsinos within the TeV scale. While the viability of
such a spectrum and its observable consequences have already been
studied~\cite{bm,misc1,misc2}, it is agreed that SUSY phenomenology in
such cases should revolve around these lighter particles. The lightest
neutralino which is the lightest SUSY particle (LSP) is in the right
mass range to become a dark matter candidate. While the electroweak
gauginos and Higgsinos can decay within a detector, heavy squarks
render the gluino necessarily long-lived. Based on cosmological
implications of a long-lived gluino, an upper bound of about $10^{13}$
GeV on the SUSY breaking scale has been suggested~\cite{ad}. In the
Higgs sector, some `fine-tuning' (which is no more an untouchable
concept) is done to keep all physical scalars excepting the lightest
neutral one heavy, and still implement electroweak symmetry breaking
at the requisite scale.
\end{itemize}

Only the fields included in the minimal SUSY standard model are
assumed to control the phenomenology in most of these studies (except
in~\cite{bm} where it is shown that some fields from the SUSY breaking
sector might have a role to play). Here we examine what happens if there
is baryon/lepton number violation incorporated in this scenario. In
other words, we explore some features of the phenomenology associated
with R-parity violation in the split SUSY model.

The MSSM superpotential, written in terms of the Higgs, quark and
lepton superfields, is given by

\begin{equation}
W_{MSSM} = h^l_{ij} L_i H_1 E^c_j + h^d_{ij} Q_i H_1 D^c_j
         + h^u_{ij} Q_i H_2 U^c_j + \mu H_1 H_2
\end{equation}

\noindent 
where $H_1$ and $H_2$ are respectively the Higgs doublets that lend
mass to the down-and up-type quarks. $i,j$ etc. are family indices.
Now, if $R=(-)^{(3B+L+2S)}$ is not conserved, then the superpotential
admits of the following additional terms~\cite{rp}:

\begin{equation}
W_{{R\slash_p}} = \lambda_{ijk} L_i L_j E^c_k 
              + \lambda^{'}_{ijk} L_i Q_j D^c_k
	      + \lambda^{''}_{ijk} \bar{U}_i D^c_j \bar{D}_k
	      + \epsilon_{i} L_{i} H_2
\end{equation}

\noindent 
where the terms proportional to $\lambda_{ijk}$, $\lambda_{ijk}^{'}$
and $\epsilon_i$ violate lepton number, and those proportional to
$\lambda_{ijk}^{''}$ violate baryon number. The constants
$\lambda_{ijk}$ ($\lambda_{ijk}^{''}$) are antisymmetric in the first
(last) two indices.  We are assuming at the beginning of this analysis
that the bilinear terms $\epsilon_{i} L_{i}H_2$ are not present. We
shall later comment on their possible implications. Also, the
simultaneous existence of the L-and B-violating terms can normally
lead to fast proton decay, and very strong constraints on the products
of such terms have been derived~\cite{pdk}. Again, we shall show at
the end how such constraints fare in the split SUSY scenario.

In general, the lightest neutralino ($\neu_1$) is unstable as a result
of the trilinear interactions. For example, we can see from figure 1
how $\neu_1$ decays into three-body final states, driven, for example, 
by the $\lambda^{'}$ type interactions~\cite{lspdk}. This usually destroys
the potential of the LSP as a dark matter candidate.

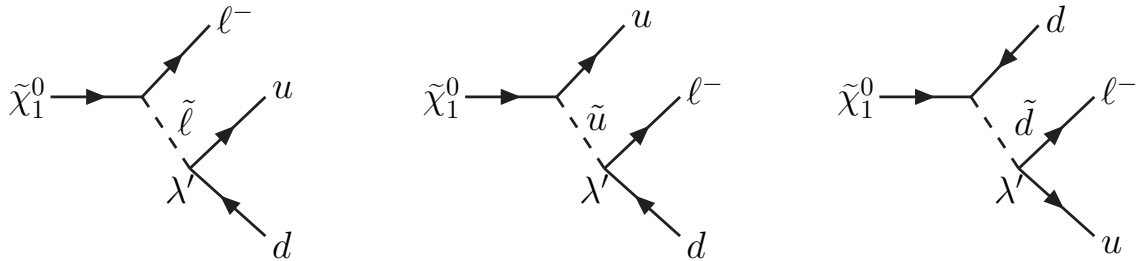
\begin{figure}[h]
{
\unitlength=1.5 pt
\SetScale{1.5}
\SetWidth{0.7}      
\normalsize  
{} \qquad\allowbreak
\begin{picture}(81,70)(24,10)
\Text(24.0,49.0)[r]{\large$\neu_1$}
\ArrowLine(25.0,49.0)(48.0,49.0)
\Text(61.0,44.0)[r]{\large$\slep$}
\DashLine(48.0,49.0)(60.0,31.0){3.0}
\Text(67.0,69.0)[l]{\large$\ell^-$}
\ArrowLine(48.0,49.0)(65.0,67.0)
\Text(81.0,51.0)[l]{\large$u$}
\ArrowLine(60.0,31.0)(79.0,49.0)
\Text(81.0,12.0)[l]{\large$d$}
\ArrowLine(79.0,14.0)(60.0,31.0)
\Text(54.0,26.0)[l]{\large$\lambda'$}
\end{picture} \
{} \qquad\allowbreak
\begin{picture}(81,70)(24,10)
\Text(24.0,49.0)[r]{\large$\neu_1$}
\ArrowLine(25.0,49.0)(48.0,49.0)
\Text(61.0,44.0)[r]{\large$\tilde{u}$}
\DashLine(48.0,49.0)(60.0,31.0){3.0}
\Text(67.0,69.0)[l]{\large$u$}
\ArrowLine(48.0,49.0)(65.0,67.0)
\Text(81.0,51.0)[l]{\large$\ell^-$}
\ArrowLine(60.0,31.0)(79.0,49.0)
\Text(81.0,12.0)[l]{\large$d$}
\ArrowLine(79.0,14.0)(60.0,31.0)
\Text(54.0,26.0)[l]{\large$\lambda'$}
\end{picture} \
{} \qquad\allowbreak
\begin{picture}(81,70)(24,10)
\Text(24.0,49.0)[r]{\large$\neu_1$}
\ArrowLine(25.0,49.0)(48.0,49.0)
\Text(64.0,44.0)[r]{\large$\tilde{d}$}
\DashLine(48.0,49.0)(60.0,31.0){3.0}
\Text(67.0,69.0)[l]{\large$d$}
\ArrowLine(65.0,67.0)(48.0,49.0)
\Text(81.0,51.0)[l]{\large$\ell^-$}
\ArrowLine(60.0,31.0)(79.0,49.0)
\Text(81.0,12.0)[l]{\large$u$}
\ArrowLine(60.0,31.0)(79.0,14.0)
\Text(54.0,26.0)[l]{\large$\lambda'$}
\end{picture} \
} 
\caption
{\footnotesize\it Representative Feynman diagrams for neutralino decay via 
$\lambda_{ijk}^{'}$ vertices.}
\end{figure}

In the case of split SUSY, however, the situation is somewhat
different.  Decays of the LSP, driven by any of the trilinear
R-violating couplings, necessarily involves squark or slepton
propagators. This causes a large suppression in the decay rate of the
LSP, and, the higher the SUSY breaking scale ($M_S$) is, the more
long-lived it will be.  Therefore, we are faced with a situation here
where lepton/baryon number violation does not prevent one from having
a SUSY dark matter candidate, at least in certain regions of the
parameter space.

The lifetime of the LSP depends on (a) the R-parity violating
coupling(s) and (b) the squark/slepton masses. It should be noted that
the experimental upper limits on the various couplings of the
$\lambda$, $\lambda^{'}$ and $\lambda^{''}$-types found in the
literature~\cite{limits} get considerably weakened and may even
disappear when the sfermions are heavy. Also, in our analysis we are
taking {\em one} coupling of a given type at a time.

Two quantities that are likely to carry the implications of a slowly
decaying LSP are (a) the lifetime of the neutralino, and (b) its decay
length. In calculating them, we have taken the typical case of an LSP
traveling with an energy of 250 GeV; for any other energy the
corresponding quantities are easily calculable using the appropriate
boost factors. Also, the properties of the LSP are controlled by the
SU(2) gaugino mass $M_2$, the Higgsino mass parameter $\mu$, and
$\tan\beta$, the ratio of the vacuum expectation values of the two
Higgs doublets. Gaugino mass unification has been assumed. We have
used $\tan\beta~=~10$. Two sets of values of the other two
parameters, namely $M_2~=200$ GeV, $\mu~=~1000$ GeV, and $M_2~=500$
GeV, $\mu~=~150$ GeV, are used to demonstrate the results. These
correspond to the gaugino-and Higgsino-dominated LSP, respectively.

Figures 2(a) and 2(b) show us the results for the $\lambda^{'}$ and
$\lambda^{''}$-type couplings. Only one coupling of a given type is
present in each case, and all decay products arising from that
particular coupling have been summed over.

\begin{figure}[h]
\setcounter{figure}{1}
\centerline{
\epsfxsize= 16.0 cm\epsfysize=12.0cm
                     \epsfbox{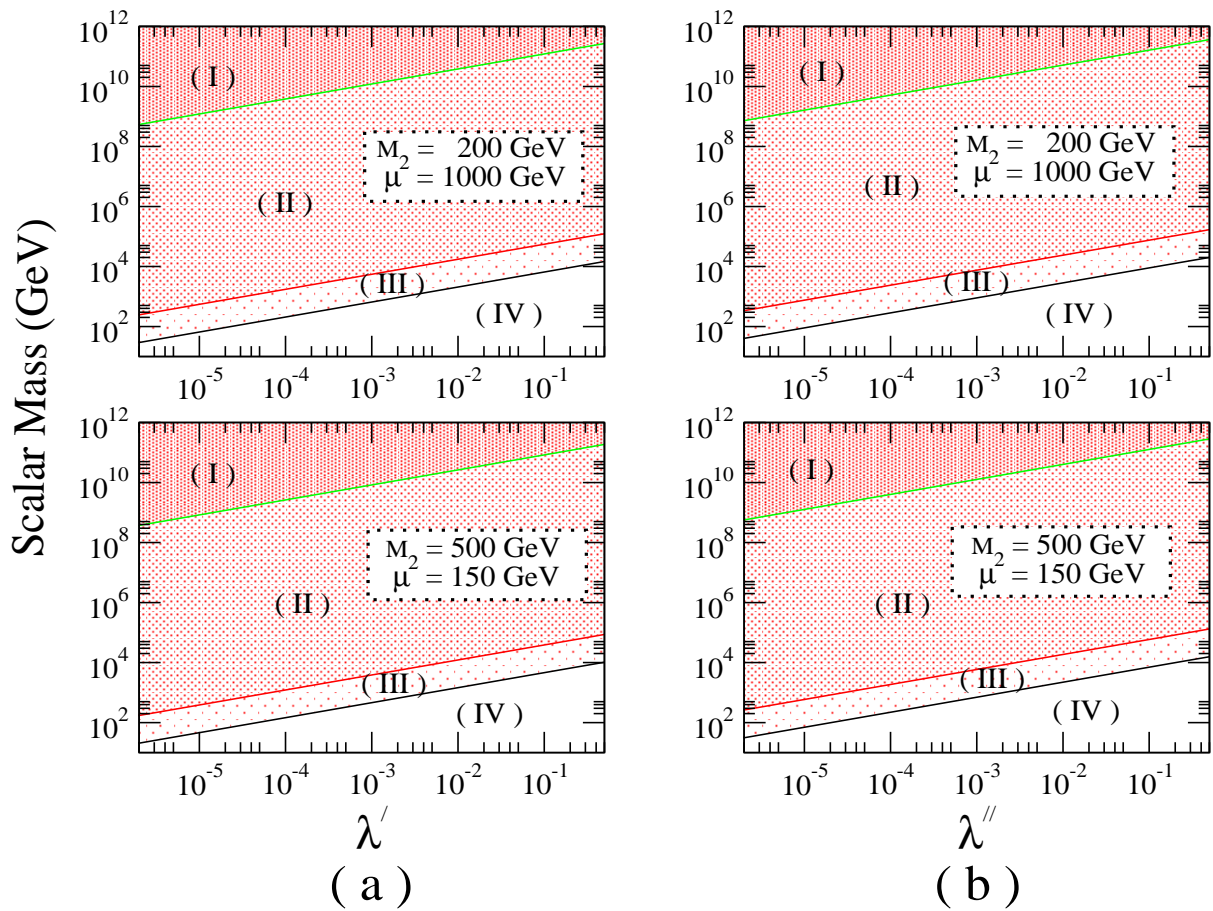}
}
\caption{\footnotesize\it
Regions corresponding to different decay rates of the lightest neutralino 
in  R-parity violating models, with (a) $\lambda^{'}$-type and
(b)  $\lambda^{''}$-type interactions. 
Both the cases for 'Bino' ($M_{2}=200~GeV, \mu=1000~GeV$) and 
'Higgsino' ($M2=500~GeV,  \mu=150~GeV$) type LSP are shown. See text 
for explanation of regions I-IV. In all cases,
neutralino energy = 250 GeV. $\tanb = 10$.}
\label{fig:23}
\end{figure}

In each case, the figures demonstrate regions of the parameter space
corresponding to the following possibilities:

\begin{enumerate}
\item The lifetime of the LSP is greater than 14 billion years,
estimated to be the age of the universe~\cite{univ}.

\item The lifetime is less than the above value, but
the decay length is more than 5 meters.

\item The decay length is more than 10 millimeters but less than 5 meters.

\item the decay length is less than 10 millimeters.
\end{enumerate}

Case (1) corresponds to region $I$ in each figure. This is seen to
happen for scalar masses (or the SUSY breaking scale) between $10^9$
and $10^{11}$ GeV, while the R-parity violating interactions varies over a
wide range. We can see from this figure that however large the L-or
B-violating interactions are, the LSP lifetime will exceed the age of
the universe, due to the massive scalar propagators. Therefore, the
lightest neutralino will continue to remain a dark matter candidate.

Case (2), represented by region $II$, is where the decay length
exceeds the order of the radius of the hadron calorimeter in, say, the
ATLAS detector at the Large Hadron Collider (LHC)~\cite{hadcal}. The
minimum scalar mass required for this varies approximately between 1
TeV and 100 TeV, as the R-parity violating coupling ranges from
$10^{-6}$ to 1. It is expected that an LSP in this region of the
parameter space has a large probability of being invisible in
accelerator experiments, in spite of R-parity violation. However,
since its average lifetime is less than the age of the universe, this
`invisible' particle still cannot be a dark matter candidate.

Region $III$ shows the region where the decay length is between 10
millimeters and 5 meters. This is a conservative description of cases
where the LSP is not only visible
but also shows a displaced vertex due to the suppression of its decay
width by the relatively large scalar masses. Depending on the value of
the coupling, this feature can be observed for the SUSY breaking scale
upto about 10 TeV or slightly above.

Finally, in region $IV$ one has an LSP of very small decay length which
makes it unstable within a very short distance scale. This is where
the usual phenomenology of weak-scale SUSY with R-parity violation
is observed. It should be
noted that the bottom right-hand corner regions of the graphs,
representing small scalar masses but large R-violating couplings, can
often be ruled out by phenomenological constraints, depending on the
particular interaction.

In order to say that region $I$ corresponds to neutralino dark matter,
one must also ensure that we have the right relic abundance arising out of 
it. Such an analysis has already been performed in the literature
{\cite{ad2},{\cite{pierce}. The contribution to relic abundance
is governed by two factors, namely, the abundance of LSP's arising
from gravitino decay/scattering, and the rate of their annihilation.
As has been shown, for example, in reference {\cite{ad2}, this translates
into a constraint on the gravitino mass if one stipulates specific values
of the scalar mass and the LSP mass. A larger LSP mass for a given scalar 
mass necessitates a higher value of the gravitino mass. it can also be
seen from the same reference that an LSP of mass $\ge 100$ GeV (i.e. in 
the range used in this paper) is 
`safe' from this standpoint so long as the gravitino mass is few times
$10^5$ GeV or above. It is also to be noted that the annihilation rate
for the (stable) neutralinos corresponding to region $I$ is not
affected by R-parity violating couplings of the trilinear type, since
no diagram driven by them leads to neutralino annihilation. For bilinear
R-parity violation, on the other hand, there is mixing between neutralinos 
and neutralinos, and thus additional  annihilation channels mediated by
the W and the Z opens up. This actually can further relax the constraints
on the parameter space of the scalar and LSP masses. However, as will be 
seen later in this paper, a bilinear R-parity violating scenario is
unlikely to provide a SUSY dark matter candidate at all.

It may also be asked whether the (late) decays of the LSP can give rise,
say, to photons at a late stage, and cause an unacceptable diffuse 
photon background.
However, this is possible if there is a substantial branching ratio
for the lightest neutralino decaying into a photon and a neutrino.
As has been shown, for example, in reference \cite{roymukh}, this 
branching ratio is very small over most of the parameter space,
especially for $m_{LSP}\ge 100$ GeV. Therefore, it is not expected
to pose any major problem. The other final states arising from LSP
decays mostly give rise to particles which are either unstable on a
cosmological scale or have high annihilation rates. However, a detailed
investigation may need to be undertaken to understand all implications
of such final states. Such an investigation is the subject of a 
separate project.

On closer scrutiny, however, region $II$ (particularly the lower part
of it) does not always correspond to an invisible LSP. For a particle
of decay length L, the probability of its decay within a distance $x$ 
is given by 

\begin{equation}
P(x)~=~(1 - e^{-x/L})
\end{equation}

\noindent
Therefore, a certain fraction of the LSP's produced must decay within
the distance characterising the size of the detector, even though it 
happens to be smaller than the decay length. The degree of visibility 
thus acquired by the LSP depends on not only the factors determining 
its decay length but also the production cross-section and luminosity in the
particular experiment.

\begin{figure}[t]
\centerline{
\epsfxsize= 10.0 cm\epsfysize=8.0cm
                     \epsfbox{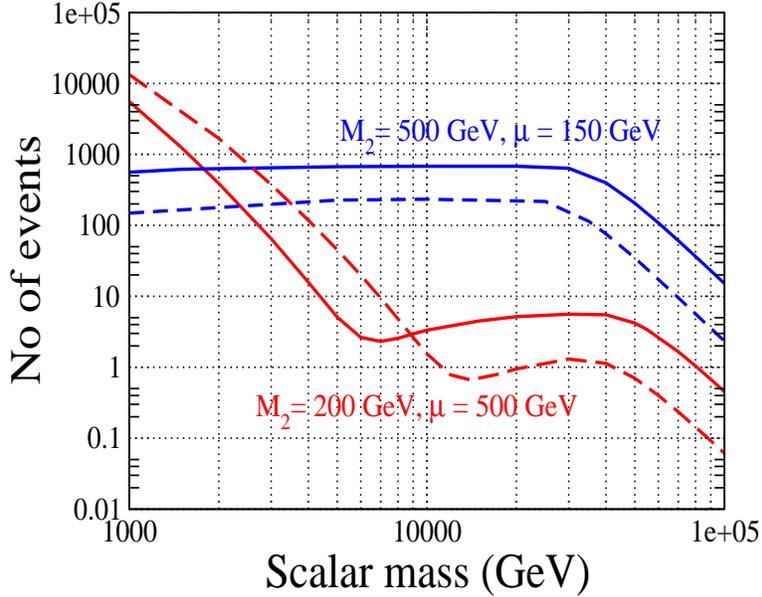}
}
\caption{\footnotesize\it
The variation of number of neutralino decay events 
detectable within the detector in a linear collider.  Solid line : CM
energy = 500 GeV, dashed line : CM energy = 1000 GeV. $\tanb = 10$. 
An integrated luminosity of 500 $fb^{-1}$ has been assumed. The 
value of the R-parity violating coupling is set at 0.1}
\label{fig:4}
\end{figure}
  
In figure 3 we show the number of LSP decay events seen within a
distance of 5 meters from the production point, for various values of
$M_S$. In order to give a general idea, we have confined ourselves to
pair-production in the process $e^{+}e^{-}\longrightarrow \neu_1
\neu_1$ at a linear electron-positron collider,
with either of the two $\chi^0_1$'s decaying within the 
specified distance. Two values of the
center-of-mass energy, namely 500 GeV and 1 TeV, are used, and the
luminosity is taken to be 500 $fb^{-1}$. For each energy, we consider
two points in the neutralino parameter space, in one of which the LSP
is gaugino-dominated, and in the other, Higgsino dominated. In the
former case, the s-channel production of the lightest neutralino pair
is more hopelessly suppressed, not only because of the usual s-channel
suppression at high energy, but also because the LSP coupling to the
Z-boson has to depend on its very small Wino content. Therefore, the
suppression of the event rate with a rise in $M_S$ is more
drastic. For a Higgsino-dominated LSP, on the other hand, there is a
stronger s-channel coupling involved, making the contributions
somewhat bigger even for relatively large values of $M_S$. Also, the
event rates presented here are without any cuts. If we
remember that in practice they are going to be further subjected to cuts
and that various detection efficiencies are involved in obtaining
the finally counted rates,
the overall conclusion is that while a gaugino-dominated LSP
indeed becomes invisible for $M_S$ exceeding a few TeV's, a Higgsino
dominated one can still exhibit a noticeable number of decay events
for the SUSY breaking scale approaching $10^5$ GeV or thereabout.

Finally, we would like to point out that the split SUSY scenario also
relaxes the constraints on R-parity violation from proton decay. Such
constraints generally lead one to rule out the simultaneous existence
of L-and B-violating terms in the superpotential.  On closer
examination, one may obtain extremely stringent limits on the products
of the two kinds of couplings, for SUSY broken within the TeV
scale. For example, one gets ~\cite{pdk}

\begin{equation}
\lambda^{'}_{11i} \lambda^{''}_{11i} \lsim  10^{-27} 
              \frac{{\tilde{m}}^2} {(100~GeV)^2}
\end{equation}

It is immediately obvious that a scenario that allows the scalar
masses to be much larger will dilute the above constraint, or cause it
to disappear altogether. For example, one can stretch the product to
values as high as, say, $10^{-11}$, for $M_S\ge 10^{10}$ GeV. This
corresponds to a case where one has simultaneous (though small) baryon
and lepton number violation and still a stable proton, while the LSP
is still invisible (with or without being a dark matter candidate).

In our discussion, we have so far left out the bilinear lepton number
violating terms $\epsilon_i L_i H_2$ in the superpotential.  Such
terms, of course, have been of considerable interest in the recent
past~\cite{bil}, in connection with a wide class of phenomena ranging
from neutrino masses to special collider signatures. It has also been
shown that even if one has just trilinear couplings at a high scale
where SUSY breaking takes place, the bilinears are radiatively induced
in the low-energy Lagrangian~\cite{barg}. In presence of the
bilinears, the scalar potential also develops additional terms,
leading to non-zero vacuum expectation values for sneutrinos. On the
whole, a general consequence is the mixing between charginos and
charged lepton on the one hand, and neutralinos and neutrinos on the
other. As a result, the lightest neutralino can decay into a neutrino
and a (real or virtual) Z or a charged lepton and a (real or
virtual)~\cite{vis} W. Since these decay rates are independent of
scalar masses, high values of $M_S$ will not suppress them, and the LSP
will decay as quickly as is originally expected. However, this is true
only if the high-scale Lagrangian itself contains the bilinears.  If
on the other hand they are induced radiatively from trilinears which
alone are present at a high scale, then these loop-induced terms are
again heavily suppressed by $M_S$, and the LSP decays driven by them
are also correspondingly suppressed. In such a situation again the
conclusions which have been drawn earlier in this paper are valid.

In conclusion, a split supersymmetry scenario can be of 
phenomenologically interesting consequence in relation to the violation 
of R-parity. They can lead to an LSP which can decay in principle but has 
a lifetime exceeding the age of the universe, being thus a dark matter
candidate still. One can also have an unstable LSP which cannot
be a dark matter candidate but still takes long enough to decay,
so that it is invisible in collider experiments. It is also possible
to have LSP decays showing displaced vertices. And finally, with
the constraints from proton decay disappearing, the simultaneous violation
of baryon and lepton numbers becomes a distinct possibility.
It may be worthwhile to probe further implications of the above 
not only in connection with laboratory experiments but also in
the cosmological context.

\newcommand{\plb}[3]{{Phys. Lett.} B #1 (#3) #2}
\newcommand{\prl}[3]{Phys. Rev. Lett. #1 (#3) #2}
\newcommand{\prep}[3]{Phys. Rep. #1 (#3) #2}
\newcommand{\rpp}[3]{Rep. Prog. Phys.  #1 (#3) #2}
\newcommand{\prd}[3]{Phys. Rev.  D #1 (#3) #2}
\newcommand{\npb}[3]{Nucl. Phys.  B #1 (#3) #2}
\newcommand{\npbps}[3]{Nucl. Phys. B (Proc. Suppl.) #1(#3)  #2}
\newcommand{\sci}[3]{Science  #1 (#3) #2}
\newcommand{\zp}[3]{Z.~Phys. C #1 (#3) #2}
\newcommand{\epj}[3]{Eur. Phys. J.  C#1 (#3) #2}
\newcommand{\mpla}[3]{Mod. Phys. Lett.  A#1 (#3) #2}
\newcommand{\ajp}[3]{{\em Am. J. Phys.\/}  #1 (#3) #2}
\newcommand{\jpg}[3]{{J. Phys.\/}  G #1 (#3) #2}
\newcommand{\astropp}[3]{Astropart. Phys.  #1 (#3) #2}
\newcommand{\ib}[3]{{ibid.\/}  #1 (#3) #2}
\newcommand{\app}[3]{{ Acta Phys. Polon.   B\/} #1 (#3) #2}
\newcommand{\nuovocim}[3]{Nuovo Cim.  C#1 (#3) #2}
\newcommand{\philt}[3]{Phil. Trans. Roy. Soc. London A  #1 (#3) #2}
\newcommand{\hepth}[1]{hep-th/#1}
\newcommand{\hepph}[1]{hep-ph/#1}
\newcommand{\hepex}[1]{hep-ex/#1}
\newcommand{\astro}[1]{astro-ph/#1}


\begin{thebibliography}{99}
\bibitem{susy}For introductory reviews see, for example,
        H. P. Nilles, \prep {110}{1}{1984}; 
	H. Haber and G. Kane, \prep {117}{75}{1985}, 
        {\it Perspectives on supersymmetry},
        G. Kane (ed),World Scientific, 1998.

\bibitem{ad} N. Arkani-Hamed and S. Dimopoulos, \hepth{0405159}.

\bibitem{cc} S. Weinberg, Rev. Mod. Phys. 61 (1989) 1.

\bibitem{lands} S. Weinberg, Phys. Rev. Lett. 59, (1987) 2607;
R. Bousso and J. Polchinski,  JHEP 0006 (2000) 006;
L. Susskind, \hepth{0302219}; M. Douglas, JHEP  0305 (2003) 046;
S. Kachru et al., \prd{68}{046005}{2003}.

\bibitem{prob} M. Douglas, \hepth{0405279}.

\bibitem{sugut} S. Dimopoulos and H. Georgi, \npb{193}{150}{1981}; 
L. Ibanez and G. Ross, \plb{105}{439}{1981};
M. Einhorn and D. Jones, \npb{196}{475}{1982}; 
U. Amaldi et al., \plb{281}{374}{1992};
P. Langacker and N. Polonsky, \prd{52}{3081}{1995}.
 
\bibitem{gut} G. Giudice and A. Romanino, \hepph{0406088}.

\bibitem{bm} B. Mukhopadhyaya and S. SenGupta, \hepth{0407225}.

\bibitem{misc1} L. Susskind, \hepth{0405189};
	A. Arvanitaki et al., \hepph{0406034};
	A. Pierce  \hepph{0406144};
	X. Calmet \hepph{0406314};
	M. Dine, E. Gorbatov and S. Thomas, \hepth{0407043};
	P. Chankowski et al., \hepph{0407242};
	E. Silverstein, \hepth{0407202};
        R. Mahbubani, \hepph{0408096}; 
	M. Binger, \hepph{0408240}.

\bibitem{misc2} S. Zhu, \hepph{0407072};
	W. Kilian et al., \hepph{0408088};
	J. Hewett et al., \hepph{0408248};
	L. Anchordoqui, H. Goldberg and Carlos Nunez, \hepph{0408284}.


\bibitem{rp} See, for example, V. Barger, G. Giudice and T. Han,
\prd{40}{2987}{1989}, see also H. Dreiner in {\it Perspective in Supersymmetry} ed., G. Kane, ({\it World Scientific}).


\bibitem{pdk} J. Goity and M. Sher, \plb{346}{69}{1995}; G. Bhattacharyya
and P. B. Pal, \prd{59}{097701}{1999}.
 
\bibitem{lspdk} H. Dreiner and P. Morawitz, \npb{428}{31}{1994};
E. A. Baltz and P. Gondolo, \hepph{9709445}.

\bibitem{limits} See, for example, B. Allanach, A. Dedes and H. Dreiner,
\prd{60}{075014}{1999}.

\bibitem{univ} L. Knox et al., \astro{0109232}; L. Krauss, 
\astro{0305556}.

\bibitem{hadcal} ATLAS technical design report, Vol.1, CERN-LHCC/99-014.

\bibitem{ad2} N. Arkani-Hamed {\it et al.}, hep-ph/0409232.

\bibitem{pierce} A. Pierce, \prd{70}{075006}{2004}.

\bibitem{roymukh} B. Mukhopadhyaya and S. Roy, \prd{60}{115012}{1999}.

\bibitem{bil} See, for example, C. S. Aulakh and R. N. Mohapatra,
\plb{119}{136}{1982};  F. de Campos et al., 
\npb{451}{3}{1995}; H. P. Nilles and N. Polonsky, \npb{484}{33}{1997}; 
S. Roy and B. Mukhopadhyaya, \prd{55}{7020}{1997}; J. Valle, \hepph{9808292};
A. Joshipura and S. Vempati, \prd{60}{111303}{1999}. For a general review see
B. Mukhopadhyaya, \hepph{0301278}.

\bibitem{barg} V. Barger et al., \prd{53}{6407}{1995}; B. de Carlos
and P. White, \prd{54}{3427}{1996}.
 

\bibitem{vis} B. Mukhopadhyaya, S. Roy and F. Vissani,  \plb{443}{191}{1998}.

\end{thebibliography}
\end{document}